\documentclass[journal=jacsat,manuscript=article]{achemso}

\usepackage{graphicx}
\usepackage{color}

\usepackage[version=3]{mhchem} % Formula subscripts using \ce{}
\usepackage[T1]{fontenc}       % Use modern font encodings

\author{Maciej Dendzik}
\author{Albert Bruix}
\author{Matteo Michiardi}
\author{Arlette S. Ngankeu}
\author{Marco Bianchi}
\author{Jill A. Miwa}
\author{Bj\o rk Hammer}
\author{Philip Hofmann}
\author{Charlotte E. Sanders}
\email{sanders.charlotte@phys.au.dk}
\phone{+45 871 55585}
\fax{+45 861 20740}
\affiliation{Department of Physics and Astronomy, Interdisciplinary Nanoscience Center, Aarhus University, 8000 Aarhus C, Denmark}

\title{Contact-Induced Semiconductor-to-Metal Transition in Single-Layer WS$_2$}

\abbreviations{IR,NMR,UV}

\begin{document}

% \affiliation{Department of Physics and Astronomy, Interdisciplinary Nanoscience Center (iNANO), Aarhus University, 8000 Aarhus C, Denmark}
% \affiliation{email: lastauthor@lastauthor.com}
%\date{\today}

% \begin{tocentry}

% \includegraphics[width=0.9\textwidth]{TOC.pdf}

% \end{tocentry}

\begin{abstract}
Low-resistance ohmic contacts are a challenge for electronic devices based on two-dimensional materials. We show that an atomically precise junction between a two-dimensional semiconductor and a metallic contact can lead to a semiconductor-to-metal transition in the two-dimensional material---a finding which points the way to a possible method of achieving low-resistance junctions. Specifically, single-layer WS$_2$ undergoes a semiconductor-to-metal transition when epitaxially grown on Ag(111), while it remains a direct band gap semiconductor on Au(111). The metallicity of the single layer on Ag(111) is established by lineshape analysis of core level photoemission spectra. Angle-resolved photoemission spectroscopy locates the metallic states near the ${Q}$ point of the WS$_2$  Brillouin zone. Density functional theory calculations show that the metallic states arise from hybridization between Ag bulk bands and the local conduction band minimum of WS$_2$ near the $Q$ point.
%This finding---combined with previous findings that substrate interactions can determine band gap size and doping in 2D semiconductors---suggests that entire electronic devices incorporating semiconductors and metallic interconnects could be fabricated by placing a single sheet of 2D material on a suitably pre-patterned substrate. 
\end{abstract}
%\pacs{\mr{79.60.Bm,73.20.At, 72.10.Fk}}
\maketitle

\section{Introduction}

Low-resistance metal-semiconductor contacts are an essential ingredient of electronic devices. This subject is well understood for bulk materials \cite{Tung:2014aa}, but is one of the biggest hurdles for the exploitation of novel single-layer (SL) semiconductors such as MoS$_2$ or WS$_2$ \cite{Allain:2015aa}. The contact resistance between a metal and a semiconductor is, in general, influenced by several factors, such as the heights of the Schottky and tunnel barriers and the degree of hybridization between the materials. These parameters cannot be inferred from the equilibrium properties of the materials in a simple way, particularly in the case of SL materials. The Schottky barrier, for instance, which depends on the SL semiconductor's electronic structure, can be affected by a gap renormalization in the presence of an underlying metal \cite{Inkson:1973aa}.

One way to achieve a low-resistance junction and avoid the occurrence of a tunneling barrier is to induce a local semiconductor-to-metal transition in the SL material itself---for example, via a local structural phase transition from the semiconducting trigonal prismatic (``1H'') polymorph to the metallic 1T polymorph  of a transition metal dichalcogenide \cite{Eda:2012aa,Kappera:2014aa,Kappera:2014ab,Cho:2015aa,Song:2016aa}.  However, from the standpoint of device applications, the practical utility of this approach has so far been limited by the difficulty of controllably patterning 1H/1T device structures within the SL, and also by the instability of the 1T phase.

Recent theoretical work has suggested an alternative approach:  a semiconductor-to-metal transition induced simply by interaction of the SL with its substrate \cite{Kang:2014aa,Allain:2015aa,Farmanbar:2016aa,Wang:2016af}. This is difficult to realize in practice because it requires a perfect interface between the substrate and the SL \cite{Gong:2013ab,Ovchinnikov:2014aa}.  Here we show that a semiconductor-to-metal transition can indeed occur in a SL semiconductor having an atomically well-defined interface with an appropriately chosen substrate. We find that SL WS$_2$ remains semiconducting on Au(111) but undergoes a semiconductor-to-metal transition on Ag(111). This transition results from the combination of the Ag(111) surface's lower work function and a hybridization between the conduction bands of WS$_2$ and Ag; together, these lead to the emergence of metallic bands with strong WS$_2$ character at the ${Q}$ point of the WS$_2$ Brillouin zone.  We directly observe the metallization of the SL via the asymmetry of the lineshape of the W 4f core level (CL) photoemission spectrum. 
 
The implications are significant, not only in that the creation of ohmic contacts is an important requirement for device applications:  The present results dovetail with recent findings that the substrate can control the size of the band gap in SL materials \cite{Ugeda:2014aa,Antonija-Grubisic-Cabo:2015aa} as well as the strength and character of doping \cite{Rosner:2016aa,Komsa:2012aa}.  Combining these effects, one could create complex two-dimensional electronic circuits, in which metallic wires are contacted to semiconducting devices with controllable $p$ or $n$ doping and tunable band gaps, all in a single SL semiconductor prepared on a suitably pre-patterned substrate material.

\section{Methods}

\subsection{Sample Preparation and Characterization}

The epitaxial growth method used here is similar to that which has been described previously for related material systems \cite{Dendzik:2015,Miwa:2015,Groenborg:2015}.  Substrates were prepared by Ne$^{+}$ sputtering and annealing in ultra-high vacuum.   W was evaporated onto the clean substrate surfaces at room temperature in an H$_2$S atmosphere, and samples were subsequently annealed to approximately 825$\sim$925~K while being continually exposed to H$_2$S.  Scanning tunneling microscopy (STM) confirmed coverage of ca. 0.7~monolayers (similar results were also seen for higher coverage up to 1.1~monolayers).  STM and low-energy electron diffraction (LEED) showed that the hexagonal atomic structure, the moir{\'e} superstructure, and the domain size of WS$_2$/Ag(111) were similar to those of WS$_2$/Au(111), which has been characterized elsewhere in detail \cite{SMAT,Dendzik:2015}.

All experiments were performed at the SGM3 beamline of the ASTRID2 synchrotron radiation source  \cite{Hoffmann:2004}.  Sample growth and measurement were carried out \textit{in situ}, without breaking vacuum. STM measurements were made at room temperature. The sample temperature of angle-resolved photoemission spectroscopy (ARPES) and LEED measurements was approximately 100~K. The energy and angular resolution of the ARPES measurements were better than 30~meV and 0.2$^{\circ}$, respectively.  X-ray photoemission spectroscopy (XPS) from shallow CLs indicated that the WS$_2$ on Ag(111) sample had negligible Se contamination of around 2\% of a monolayer.

\subsection{Theoretical Methods}

The electronic structure calculations were carried out using the periodic density functional theory code VASP \cite{Kresse:1993v1,Kresse:1996v2,Kresse:1996v3}. The valence electrons were described with plane-wave basis sets with a kinetic energy threshold of 415\,eV, and the interaction between the valence and frozen core-electrons was accounted for by means of the projector-augmented-wave method of Bl\"ochl \cite{Blochl:1994}. The PBE approximation to the exchange-correlation functional was used \cite{Perdew:1996} in combination with the DFT-D3 method of Grimme correction to account for vdW interactions. 

The supercell models consist of a $\sqrt{13}\times\sqrt{13}$ R13.9$^{\circ}$ cell of WS$_2$ on a 6-layer 4$\times$4 cell of the metal (111) surface. This model is smaller than the experimentally identified moir{\'e} lattice, but it allows solving the inherent mismatch between WS$_2$ and metal lattices while applying a minimal strain on the metal ($<$0.2\%). The geometry of the supercells was optimized until the forces on all atoms were smaller than 0.01\,eV\r{A}$^{-1}$. All atoms were relaxed except those in the 4 lowermost metal layers, which were kept in their truncated bulk positions. A (4$\times$4$\times$1) mesh of k-points was used to sample the reciprocal space during geometry optimizations, and the charge density was subsequently recalculated with a single point calculation using a denser (10$\times$10$\times$1) mesh of $k$-points. An energy threshold of $10^{-6}$\,eV was used to define convergence of the self-consistent field of the electron density.

For the supercell models, the band structure along the high symmetry directions of the primitive cell of WS$_2$ is folded into the smaller reciprocal lattice of the supercell. In order to recover the unfolded band structure with the symmetry of the primitive unit cell of WS$_2$, we have calculated the effective band structure using the method proposed by Popescu and Zunger \cite{Popescu:2012} as implemented in the BandUp code \cite{Medeiros:2014,Medeiros:2015aa}. Spin-orbit coupling has been included for all band structure calculations using the virtual crystal approximation as implemented in the VASP code \cite{Steiner:2016aa}.

\section{Results and Discussion}

\begin{figure}
	\includegraphics[width=0.8\textwidth]{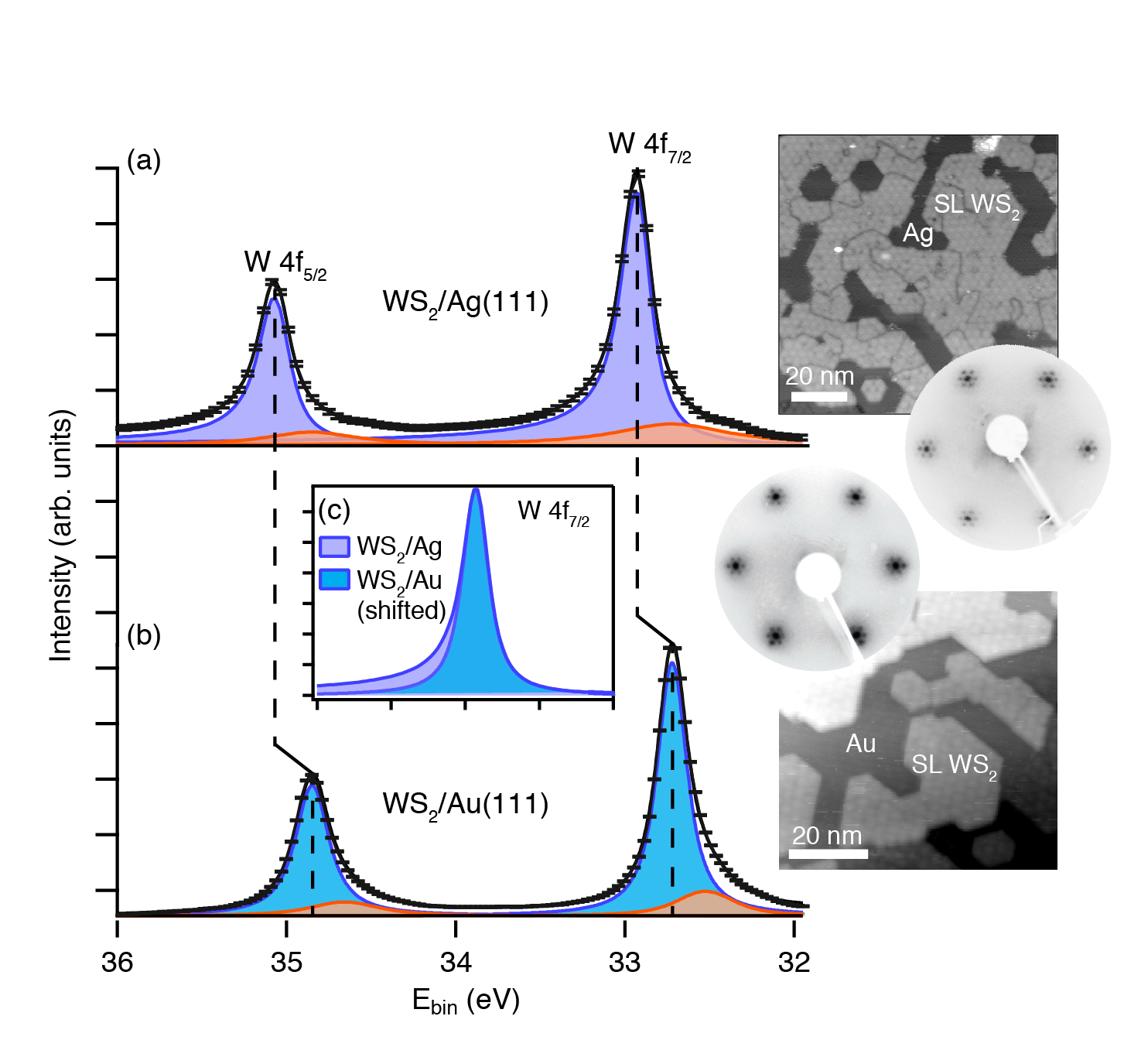}\\
	\caption{Direct evidence of metallic SL WS$_2$ on Ag(111) from CL spectroscopy. (a)--(b)  CL spectra (acquired at photon energy $h \nu$=140~eV) from the W 4f states of (a) WS$_2$/Ag(111) and (b) WS$_2$/Au(111). Data points are black crosses; the black line is the fit; blue and orange peaks correspond to WS$_2$ and incompletely sulphided WS$_{2-x}$ clusters, respectively \cite{Bruix:2015}. (c) Comparison of the fitted W 4f$_{7/2}$ components for WS$_2$/Ag(111) and WS$_2$/Au(111), shifted and normalized so that the peak maxima coincide. The insets show STM images and LEED patterns of WS$_2$ grown on Ag(111) and Au(111), respectively \cite{Horcas:2007}.  STM imaging parameters are  0.09~nA and 1.43~V in (a) and 0.58~nA and 1.16~V in (b). The electron kinetic energy for LEED is 114~eV.}
	\label{fig:CL}
\end{figure}

STM and LEED measurements show close structural similarity between WS$_2$ grown on Au(111) and on Ag(111), including the mori\'e pattern formed between SL and substrate, as seen in the insets of Fig. \ref{fig:CL} and in Refs. \cite{Dendzik:2015,Ulstrup:2016,SMAT}. However, XPS reveals that the two systems have entirely different electronic characters: This manifests itself in the shape of the W 4f CL spectra. To understand this difference, one must first notice that in Fig. \ref{fig:CL} the spectra obtained from samples grown on each of the two types of substrates (Au and Ag) consist of the 7/2-5/2 doublet, each peak of which is itself fittable by two components, one of higher intensity and one of lower intensity.  The higher-intensity component arises from the WS$_2$ basal plane, and the lower-intensity component from edge atoms or partially-sulphided WS$_{2-x}$ clusters \cite{Fuchtbauer:2013,Bruix:2015}.

The key difference between the spectra from samples prepared on each of the two substrates is the lineshapes of the high-intensity peaks.  For WS$_2$/Au(111), these can be fitted by Gaussian-broadened Lorentzians, as expected for CLs of semiconducting materials, and as has also been observed for WS$_2$ on transition metal oxides \cite{Ulstrup:2016ac}. By contrast, the peak shapes for WS$_2$/Ag(111) cannot be decomposed into a small number of Lorentzians, because of their asymmetric lineshape consisting of a tail at high binding energy.  This is particularly clear in Fig. \ref{fig:CL}(c), which shows the high-intensity components of the two systems superimposed and shifted so that the peak maxima coincide. The asymmetric lineshape is fit best by a Doniach-{\v S}unji{\'c} profile, a profile in which a Gaussian-broadened Lorentzian is modified by an asymmetry parameter that describes low-energy electron-hole interactions during the photoemission process \cite{Doniach:1970, Hufner:2003}.  While many factors can influence the shape of the CL peak, the particular asymmetry embodied by the Doniach-{\v S}unji{\'c} lineshape is a definite signature of metallicity.  (The reverse is not necessarily true:  not all metals exhibit an asymmetrical peak shape.)  Fitting of the CLs of WS$_2$/Ag(111) (described in detail in the supplementary materials) gives an asymmetry parameter of 0.10(1) for the high-intensity components, consistent with the asymmetry of other, typical metallic systems \cite{Citrin:1977aa,SMAT}.  Thus, although WS$_2$/Au(111) may retain its semiconducting character, WS$_2$/Ag(111) is metallic. We note that neither a presence of the metallic 1T phase nor edge states can be responsible for the asymmetric line shape observed here, since both have characteristic CL peaks at \emph{lower} binding energies  \cite{Bruix:2015,Bruix:2016f,Eda:2011,Acerce:2015}. Moreover, there are no signatures of the 1T phase's distinctively different band structure in the ARPES data, and significant contributions from edge states seem unlikely since the STM data reveals the formation of large islands.

An additional experimental observation relating to the metallicity of WS$_2$/Ag(111) lies in the absolute binding energy of the 4f CLs:  Specifically, the CL WS$_2$/Ag(111) peaks are shifted toward higher binding energy than those of WS$_2$/Au(111) by approximately 207(20)~meV, whereas the corresponding shift of the valence band states is significantly larger (275(24)~meV---see below). As already mentioned, the different work functions of Ag(111) and Au(111) lead to different Fermi level pinning in the two systems, explaining a shift of the valence band. However, the different shift in the CL binding energies relative to the valence states in the two systems suggests a mechanism beyond a purely electrostatic shift. A plausible explanation for this difference is local final state screening that affects the core level binding energies. In a metallic system, the core hole can be screened by the electrons at the Fermi level, leading to higher kinetic energy of the escaping photoelectron. These screening effects depend crucially on the local electronic structure around the emitting atom  \cite{Lizzit:1998aa,Bruix:2015}. The observation of a decreased binding energy / increased kinetic energy for the core electrons from WS$_2$/Ag(111) is thus consistent with a metallic SL.

\begin{figure*}
	\includegraphics[width=0.8\textwidth]{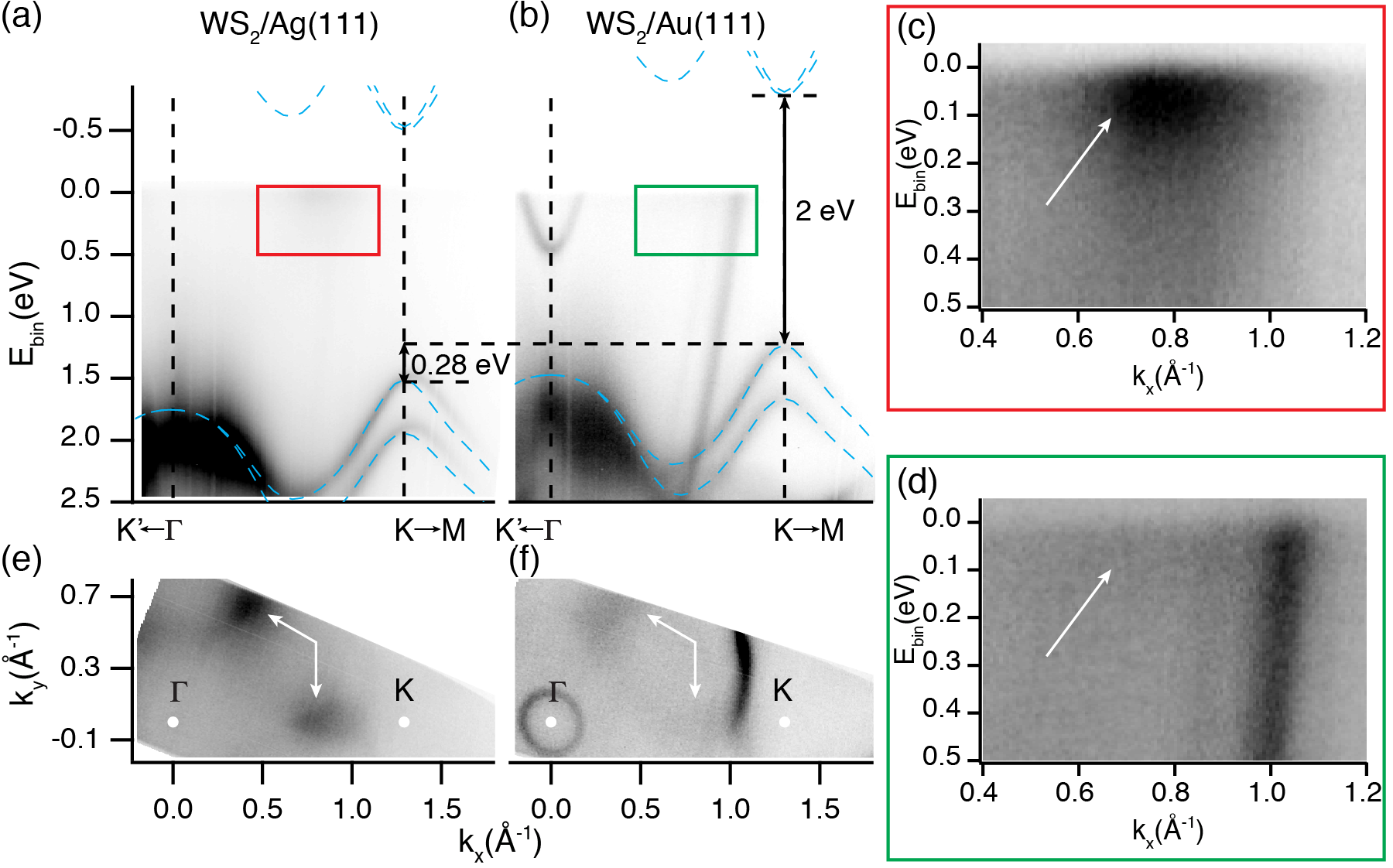}\\
	\caption{Electronic structure of SL WS$_2$ on Ag(111) and Au(111) from ARPES, with evidence for metallic states in the former system. (a)--(b) Photoemission intensity along the high-symmetry direction ${\Gamma}$-${K}$ for (a) SL WS$_2$/Ag(111) and (b) SL WS$_2$/Au(111) (coverage ca. 0.7 ML in both cases, acquired at photon energy $h \nu$=25~eV; note that the Au bulk band structure is more strongly visible at this photon energy than the Ag band structure). The blue dashed lines are calculated bands for the highest VB and lowest conduction band of free-standing SL WS$_2$, shifted such that the VB maximum coincides with the data measured here, and the conduction band minimum with the result of  pump/probe time-resolved ARPES for SL WS$_2$/Ag(111) \cite{Ulstrup:2016}.  We assume the same band gap for SL WS$_2$/Au(111). (c) and (d) show magnifications of the regions in the coloured rectangles of (a) and (b), respectively, illustrating the additional intensity near the Fermi level for WS$_2$/Ag(111), marked by a white arrow. (e)--(f) Photoemission intensity  at the Fermi energy for WS$_2$/Ag(111) and WS$_2$/Au(111) ($h \nu$=25~eV). The enhanced diffuse photoemission intensity seen in (a) and (c) between ${\Gamma}$ and ${K}$ is visible in (e) and marked by white arrows. These states are almost absent for WS$_2$/Au(111) and not found on clean Ag(111)~\cite{SMAT}.}
	\label{fig:ARPES}
\end{figure*}

The metallic character of WS$_2$/Ag(111) should be identifiable in the electronic band structure as measured by ARPES. Fig. \ref{fig:ARPES}(a) and (b) show the photoemission intensity resulting from such a measurement along the ${\Gamma}$-${K}$ direction of the Brillouin zone for WS$_2$ grown on (a) Ag(111)  and (b) Au(111). In both cases, the upper valence band (VB) of WS$_2$  with the spin-split VB maximum at ${K}$ is clearly identified and the dispersion is very similar to that calculated for a free-standing SL, shifted such that the VB maximum coincides with the measurement (blue dashed line) \cite{Dendzik:2015,SMAT}. A distortion of the bands near ${\Gamma}$ is consistent with previous observations of SL MoS$_2$/Au(111) and SL WS$_2$/Au(111)  \cite{Dendzik:2015,Miwa:2015,Bruix:2016}, and will be discussed below. The VB maximum in WS$_2$/Ag(111) is at a higher binding energy than in WS$_2$/Au(111) (by ca. 0.28~eV).

The metallic behaviour of WS$_2$/Ag(111) has to be caused by additional states near the Fermi energy and these should give rise to ARPES features that cannot be attributed to the substrate.  Indeed, as can be seen in Fig. \ref{fig:ARPES}(a), (c) and (e), there is a diffuse feature at the Fermi level along ${\Gamma}$-${K}$, marked by white arrows. These states are absent for clean Ag scanned at the same photon energy \cite{SMAT}.  Faint states can also be seen in this location in the data set acquired from WS$_2$/Au(111), but they are extremely weak in that case.  We can therefore conclude that similar physical mechanisms are at work in both systems, and that the difference in the density of states (DOS) at the Fermi level between the two systems is a quantitative effect rather than a qualitative one.  However, the effect is so weak in the case of WS$_2$/Au(111) that the system essentially retains its semiconducting character and shows no detectable sign of metallicity in CL spectroscopy (Fig. \ref{fig:ARPES}(b), (d) and (f)).

We now discuss several possible scenarios to explain these findings. The obvious interpretation is that of a band crossing from a state just above the Fermi level, such that a shallow electron pocket is formed. The simplest scenario to achieve this would be a rigid shift of the conduction band (CB) to lower energies. The calculated lowest CB for free-standing SL WS$_2$ is superimposed on the data in Fig. \ref{fig:ARPES}(a), aligning the dispersion such that the CB minimum at $K$ is found at the same energy as recently reported from a time-resolved pump/probe ARPES experiment for the same system \cite{Ulstrup:2016}. Assuming that the two material systems (WS$_2$/Ag(111) and WS$_2$/Au(111)) exhibit a similar size of the renormalized band gap, the CB is expected to be closer to the Fermi energy for WS$_2$/Ag(111) than for WS$_2$/Au(111) because of the different work function of the two metal surfaces, leading to the observed shift in the VB. While the shifted CB does indeed have a \emph{local} minimum along ${\Gamma}$-${K}$, at the so-called ${Q}$ point, this state cannot give rise to the observed Fermi level crossing in a rigid band shift picture because the \emph{absolute} CB minimum is found at ${K}$, not at $Q$, leaving the local minimum at ${Q}$ well above the Fermi level. 

Another possibility is that a distortion might occur in the lowest-lying CB, so as to pull the local minimum at ${Q}$ down to the Fermi level while leaving the feature at ${K}$ at a higher energy. Such band distortion could have several causes. Previous theoretical work has predicted this very effect as a result of in-plane compressive strain as small as 1\% \cite{Amin:2014}. However, analysis of our LEED data rules out strain larger than $\pm$0.7\% \cite{SMAT}.  Moreover, the lattice constant of Ag is larger than that of Au (4.09~\AA~for Ag versus 4.08~\AA~for Au), so it is unlikely that the WS$_2$ overlayer---if it is compressed---is more compressed on Ag than on Au.  

The most likely explanation is that the distortion of the WS$_2$ CB that results in the metallicity of the layer is caused by hybridization with the underlying Ag states. Such hybridization with substrate states has recently been shown to be responsible for the aforementioned shift of the valence states at ${\Gamma}$ toward higher binding energy than in the  freestanding SL (see comparison between the calculated VB for a free standing layer and the actually observed dispersion in Fig. \ref{fig:ARPES}(a) and (b))~\cite{Bruix:2016}. Additionally, hybridization has been suggested as being responsible for important effects in other, related two-dimensional (2D) material systems; for example, as a source of ``pseudo-doping'' in metallic SL TaS$_2$ on Au(111) \cite{Sanders:2016aa,Wehling:2016aa}.

\begin{figure*}
   \includegraphics[width=0.5\textwidth]{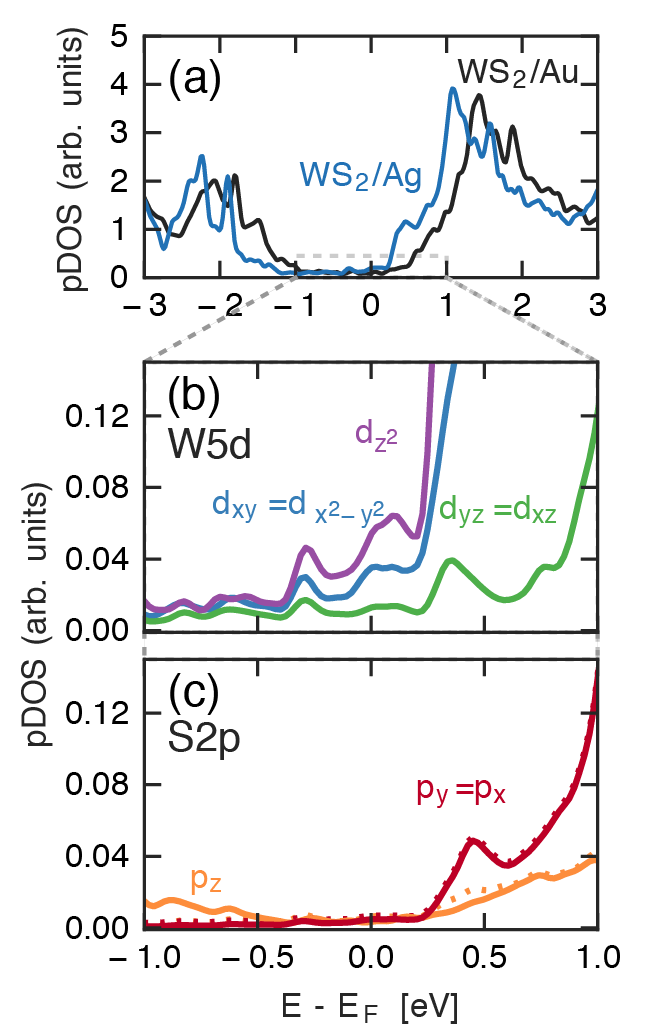}\\
   \caption{Origin of the metallic states in SL WS$_2$ on Ag(111) from hybridization, viewed in terms of the density of states. (a) Calculated densities of states projected onto the WS$_2$ states of Ag(111) and Au(111)  (pDOS). (b)--(c) Decomposed pDOS for W 5d and S 2p orbitals.}
   \label{fig:Theory_DOS}
\end{figure*}

We explore the possibility of hybridization in more detail using density functional theory calculations of SL WS$_2$ adsorbed on Ag(111) and Au(111).  The resulting densities of states projected onto WS$_2$ (pDOS) for both substrates are shown in Fig. \ref{fig:Theory_DOS}(a). The overall shift of the WS$_2$ states by 270~meV toward lower energies for WS$_2$/Ag(111) relative to WS$_2$/Au(111) agrees with the experimental binding energy shifts of Fig. \ref{fig:ARPES}. It is consistent with the lower work function of Ag, and the shift also leads to a smaller energy barrier between the metal Fermi level and the CB minimum of WS$_2$. Moreover, the calculated pDOS at the Fermi level of WS$_2$/Ag(111) is indeed non-zero, due to the CB tailing into the band gap. This supports the observation of the metallicity of the WS$_2$ layer.  We note that the conduction band actually tails into the band gap in both systems; it is the difference in relative distance between the Fermi level and the bottom of the CB that allows for a larger DOS at the Fermi level in the case of WS$_2$/Ag(111).

We further elucidate the origin of the states at the Fermi level by decomposing the pDOS into contributions from the W 5d and the S 2p orbitals (Fig. \ref{fig:Theory_DOS}(b),(c)). This reveals that the most significant contribution to the pDOS at the Fermi level stems from W d$_{xy}$, W d$_{x^2-y^2}$, and W d$_{z^2}$ orbitals. These orbitals are also the ones giving the strongest contribution to the CB near ${Q}$ for the free-standing WS$_2$ layer (see Figs. S3  and S4 \cite{SMAT} and ref. \cite{Cappelluti:2013aa}), suggesting that hybridization of the WS$_2$ CB with Ag states at this region of $k$-space is indeed responsible for the metallization of WS$_2$. 

\begin{figure*}
   \includegraphics[width=0.8\textwidth]{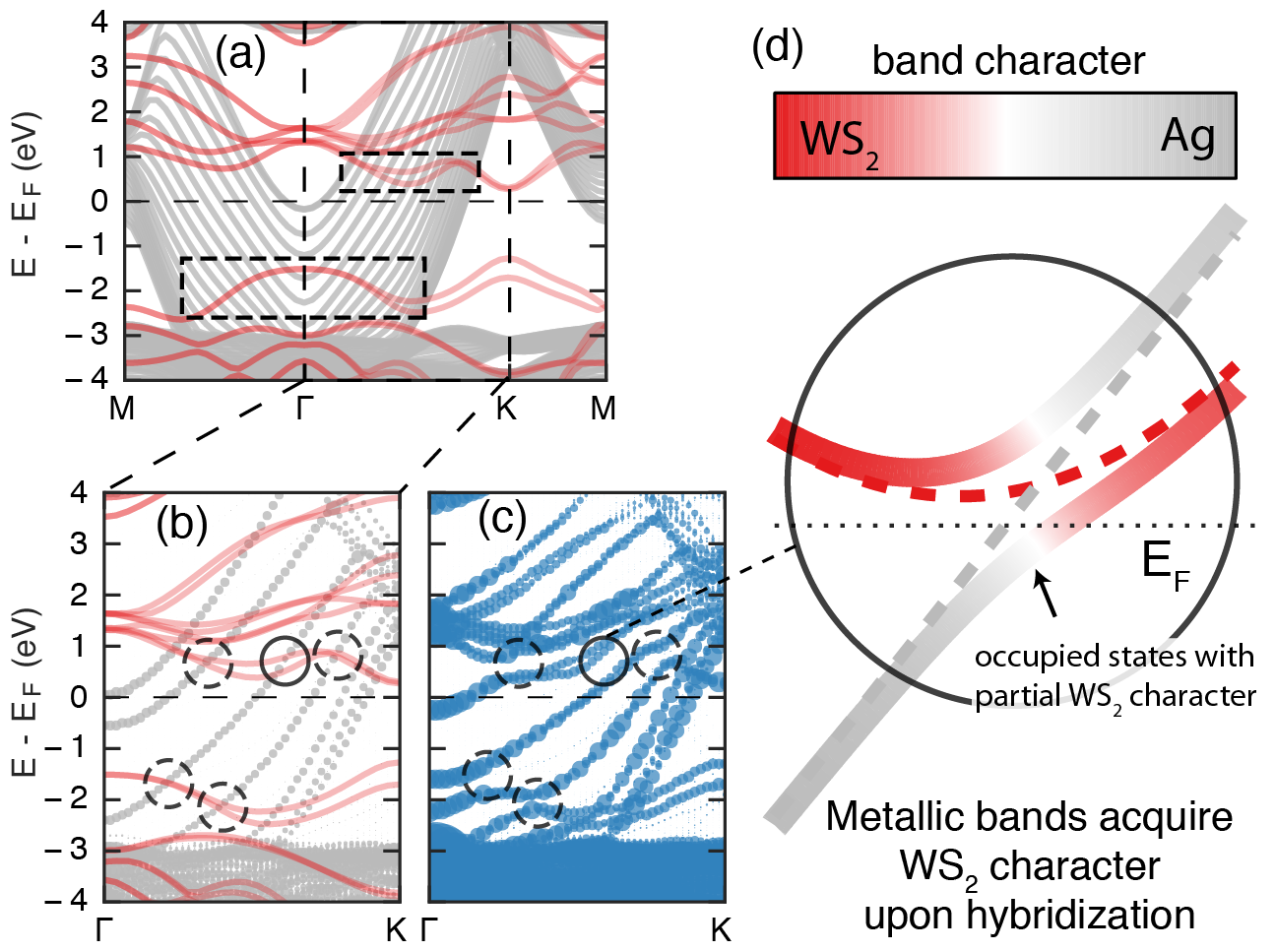}\\
   \caption{Origin of the metallic states in SL WS$_2$ on Ag(111) from hybridization, viewed in a band structure picture. (a) Calculated band structure for free-standing SL WS$_2$ (red) together with the band structure of bare Ag(111) (gray). Rectangles indicate the regions where Ag bands overlap with the valence and conduction bands of WS$_2$. (b) Calculated band structure for free-standing SL WS$_2$ (red) together with the effective  band structure of a 4$\times$4 cell of Ag(111). (c) Effective band structure of the supercell consisting of a $\sqrt{13}\times\sqrt{13}$ R13.9$^{\circ}$ of WS$_2$ on a 4$\times$4 cell of Ag(111). (d) Band hybridization leading to avoided crossing and partial character changes, as calculated from a simple two-state model. Thin dotted lines represent non-hybridized WS$_2$ and Ag bands. Thick solid lines represent the resulting hybridized bands and are colored according to their character as indicated in the overlying colormap. The solid circle in (b) and (c) corresponds to the situation depicted in (d).}
   \label{fig:Theory_bands}
\end{figure*}

This is confirmed by band structure calculations for the interacting system shown in Fig. \ref{fig:Theory_bands}. We start by plotting  the band structure of free-standing WS$_2$ (red lines) together with that of a 20-layer surface model of bare Ag(111) (gray lines) in  Fig. \ref{fig:Theory_bands}(a) and do indeed find that regions of significant overlap between WS$_2$ and Ag bands (highlighted with black rectangles) correspond to $k$-vectors near ${\Gamma}$ and ${Q}$ for the VB and CB of SL WS$_2$, respectively. To further clarify the effect of hybridization of the interacting bands, we have calculated the effective band structure of WS$_2$/Ag(111)---\textit{i.e.,} the effective weight of the bands in the large supercell when projected onto the $1 \times 1$ unit cell of the SL WS$_2$.  Fig. \ref{fig:Theory_bands}(b) shows the result of a calculation without SL substrate interaction for a bare 6-layer 4$\times$4 Ag(111) supercell together with the band structure of free-standing SL WS$_2$, and  Fig. \ref{fig:Theory_bands}(c) shows the effective band structure for the same system but including interactions.   The smaller metal slab thickness used here leads to fewer bands crossing the Fermi level than for the 20-layer 1$\times$1 unit cell shown in Fig. \ref{fig:Theory_bands}(a). Nevertheless, also here we identify crossings between WS$_2$ and Ag bands, which are indicated with black circles in  Fig. \ref{fig:Theory_bands}(b). In the effective band structure in Fig. \ref{fig:Theory_bands}(c), these crossings are avoided due to band hybridization, which also results in Ag and WS$_2$ bands acquiring partial WS$_2$ and Ag character, respectively. 

For clarity, the effect of hybridization on the Ag and WS$_2$ bands is illustrated in  Fig. \ref{fig:Theory_bands}(d), which shows the changes in band dispersion using a simple model with two interacting bands. The avoided crossing transforms one unoccupied band of WS$_2$ and one metallic band of Ag into two new bands (one unoccupied and one metallic) with varying Ag and WS$_2$ character. This effectively leads to an up-shift of the CBM of the unoccupied band, which is thus farther from the Fermi level and does not contribute to the metallicity of WS$_2$. Instead, the formation of a hybridized metallic band gives rise to occupied states with partial WS$_2$ character, resulting in the metallic character of WS$_2$. The occupied states of the metallic bands acquire the strongest contribution of WS$_2$ at ${k}$-vectors where the energy difference between the CB of WS$_2$ and the occupied Ag bands is smallest. This explains the dominant role of the CB near ${Q}$ in the semiconductor-to-metal transition:  the local CB minimum interacts with the Ag states, while the states at $K$, which constitute the absolute minimum in the CB of the noninteracting SL, do not. Since the CB of WS$_2$ is much closer to the Fermi level of Ag(111) than of Au(111), the hybridization mechanism can induce a semiconductor-to-metal transition only in the former case.

\section{Conclusions}

The results reported here show that metallization of a SL semiconductor can be achieved through substrate interaction.  We emphasize that this effect depends on high interfacial quality, and would not be expected---at least in the material systems examined here---in samples prepared \textit{ex situ} by means of, for example, ``Scotch tape'' style preparation.  It would also not necessarily be expected to occur underneath polycrystalline metallic contacts evaporated thermally onto the top of SL devices, as in traditional device architectures.   However, the fabrication is achieved by a straightforward, well-developed epitaxial approach that is easily adaptable for scalable production.  Our results are achieved in WS$_2$ on Ag(111), but there are presumably other material systems in which strong hybridization occurs between the electronic states of the SL and the substrate and in which similar (or even more dramatic) effects can be observed; identifying such systems remains a challenge for future work.

The finding that substrate interactions alone are sufficient to effect a transition from a semiconductor to a metal clears the path for major advances in 2D device fabrication. This result, in combination with earlier results showing the power of substrate interaction to tune band gap \cite{Ugeda:2014aa,Antonija-Grubisic-Cabo:2015aa} and doping \cite{Rosner:2016aa,Komsa:2012aa}, reduces the challenge of creating complex 2D device architectures from one of pattering the SL directly---a major challenge to date---to one of substrate choice and patterning, prior to covering the substrate with a homogenous SL semiconductor.

%%%

\providecommand*\mcitethebibliography{\thebibliography}
\csname @ifundefined\endcsname{endmcitethebibliography}
  {\let\endmcitethebibliography\endthebibliography}{}

%\bibliographystyle{apsrev}
%\bibliography{WS2Ag}
	%{}

\section*{ACKNOWLEDGMENTS}
We gratefully acknowledge experimental help by Fabian Arnold and stimulating discussions with Silvano Lizzit. This work was supported by the Danish Council for Independent Research, Natural Sciences under the Sapere Aude program (Grants No. DFF-4002-00029 and No. 0602-02566B) and by VILLUM FONDEN via the Centre of Excellence for Dirac Materials (Grant No. 11744). A.B. acknowledges support from the European Research Council under the European Union's Seventh Framework Programme (FP 2007-2013)/Marie Curie Actions/Grant No. 626764 (Nano-DeSign).

\end{document}